\documentclass[aps,pra,10pt,
               preprintnumbers,numbers,sort&compress,nofootinbib,
               twocolumn,%superscriptaddress,
               floatfix
               ]{revtex4}
\usepackage{graphicx,amsmath,amssymb,bm}
\usepackage{psfrag}
\usepackage{hyperref}
\usepackage{dcolumn}

\newcommand{\DFT}{\textsc{dft}}
\newcommand{\MIT}{\textsc{mit}}

\newcommand{\SLDA}{\textsc{slda}}
\newcommand{\SF}{\textsc{sf}}
\newcommand{\MC}{\textsc{mc}}
\newcommand{\LOFF}{\textsc{loff}}
\newcommand{\LO}{\textsc{lo}}
\newcommand{\ASLDA}{\textsc{aslda}}

\newcommand{\BCS}{\textsc{bcs}}

\newcommand{\vect}[1]{{\bm #1}}
\newcommand{\op}[1]{{#1}}
\renewcommand{\d}{\mathrm{d}}
\newcommand{\abs}[1]{|#1|}
\newcommand{\pdiff}[2]{\frac{\partial{#1}}{\partial{#2}}}
\begin{document}

\title{The Asymmetric Superfluid Local Density Approximation (ASLDA)}
\author{Aurel Bulgac}
%\email[E-mail:~]{bulgac@phys.washington.edu}
\author{Michael McNeil Forbes}
%\email[E-mail:~]{mforbes@alum.mit.edu}
\affiliation{Department of Physics, University of Washington,
Seattle, WA 98195-1560}
\date{\today}
%\keywords{Asymmetric Fermi gases, superfluidity, thermodynamics,
%  phase structure, density profiles, phase separation, cold atoms}
\begin{abstract}
  Here we describe the form of the Asymmetric Superfluid Local Density
  Approximation (\ASLDA), a Density Functional Theory (\DFT) used to
  model the two-component unitary Fermi gas.  We give the rational
  behind the functional, and describe explicitly how we determine the
  form of the \DFT\ from the to the available numerical and
  experimental data.
\end{abstract}
\maketitle
%\tableofcontents
\section{Introduction}
Here we describe the formulation of the Asymmetric Superfluid Local
Density Approximation (\ASLDA), which is a Density Functional Theory
(\DFT) describing normal and superfluid systems comprising two species
of fermion.  We show how the Monte-Carlo data of
Refs.~\cite{CCPS:2003,Carlson:2005kg,Carlson:private_comm2,LRGS:2006,CRLC:2007,PS:2008,BDMW:2008,BDM:2008}
was incorporated into the \ASLDA\ \DFT\ used in the papers
\cite{Bulgac:2007a,BF:2008}.  In its present formulation, the \ASLDA\
functional describes two species of fermions -- denoted $a$ (spin-up)
and $b$ (spin-down) -- with equal masses $m_{a} = m_{b} = m$, and
interacting through a resonant attractive inter-species
\textit{s}-wave two-body interaction described solely by the infinite
two-body scattering length $a_{s} = \infty$.  Throughout this paper,
we use units where $\hbar = m = 1$ to simplify the notation.

\begin{figure}[bt]
  \begin{center}
    \psfrag{x=n_b/n_a}{$x=n_{b}/n_{a}$}
    \psfrag{g(x)}{$g(x)$}
    \psfrag{x (radians)}{$x$ (radians)}
    \psfrag{y}{$y$}
    \psfrag{0.9}{\footnotesize 0.9}
    \psfrag{1.0}{\footnotesize 1.0}
    \psfrag{1.1}{\footnotesize 1.1}
    \psfrag{0.0}{\footnotesize 0.0}
    \psfrag{0.2}{\footnotesize 0.2}
    \psfrag{0.4}{\footnotesize 0.4}
    \psfrag{0.6}{\footnotesize 0.6}
    \psfrag{0.8}{\footnotesize 0.8}
    \includegraphics[width=\columnwidth,height=1.5in]{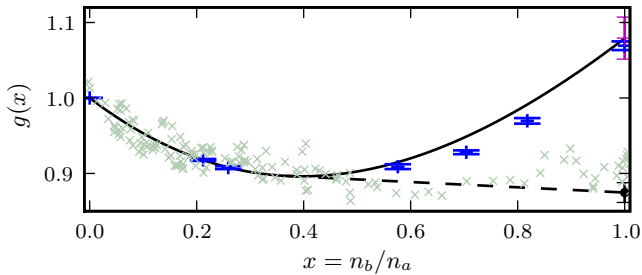}
    \caption{\label{fig:giorgini} (Color online) The dimensionless
      convex function $g(x)$ \cite{Bulgac:2006cv} that defines the
      energy density $\mathcal{E}(n_a,n_b) =
      \tfrac{3}{5}\tfrac{\hbar^2}{2m}(6\pi^2)^{2/3}\left[n_{a}
        g(x)\right]^{5/3}$. The points with error-bars (blue online)
      are the Monte-Carlo data from
      Refs.~\cite{LRGS:2006,CRLC:2007,PS:2008}.  The
      fully-paired solution $g(1)=(2\xi)^{3/5}$ is indicated to the
      bottom right, and the recent \MIT\ data~\cite{shin-2008} is
      shown (light $\times$) for comparison.  The phase separation
      discussed in Refs.~\cite{LRGS:2006,CRLC:2007,PS:2008} is
      shown by the Maxwell construction (thin black dashed line).}
  \end{center}
\end{figure}
\begin{figure}[bp]
  \begin{center}
    \psfrag{k2/kF2}{$k^2/k_{F}^2$}
    \psfrag{Eqp/eF}{$E_{k}/\epsilon_{F}$}
    \psfrag{0.0}{\footnotesize 0.0}
    \psfrag{0.4}{\footnotesize 0.4}
    \psfrag{0.8}{\footnotesize 0.8}
    \psfrag{1.2}{\footnotesize 1.2}
    \psfrag{0.5}{\footnotesize 0.5}
    \psfrag{0.7}{\footnotesize 0.7}
    \psfrag{0.9}{\footnotesize 0.9}
    \psfrag{1.1}{\footnotesize 1.1}
    \includegraphics[width=\columnwidth]{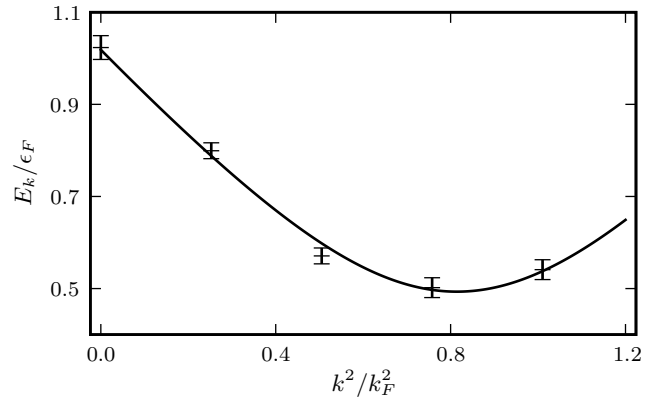}
    \caption{\label{fig:qp1} Fit of the Monte-Carlo data for the
      quasiparticle dispersions from~\cite{Carlson:2005kg} with the
      \BCS\ form~(\ref{eq:Eqp}).  This, along with $\xi$ is used to
      determine the parameters of the fully paired superfluid phase
      \SF.}
  \end{center}
\end{figure}

The model is tuned to reproduce the thermodynamics of the homogeneous
normal phase, and the homogeneous symmetric ($n_{a}=n_{b}$) superfluid
phase (\SF).  Properties of these phases have been evaluated using
accurate non-perturbative Monte-Carlo calculations, including the
energy of the interacting normal state with varying degrees of
polarization $n_{a} \neq n_{b}$ (Fig.~\ref{fig:giorgini}), the energy
of the \SF\ phase $\xi = \mathcal{E}/\mathcal{E}_{FG} =0.40(1)$
\cite{Carlson:2005kg,Carlson:private_comm2}, the quasi-particle
dispersion relationship in the \SF\ phase
(Fig.~\ref{fig:qp1})~\cite{Carlson:2005kg}, and the effective mass of
a spin-down fermion immersed in a sea of spin-up fermions
\cite{LRGS:2006,CRLC:2007,PS:2008}. We describe here how
to incorporate this high-quality information into the \ASLDA\
functional, allowing it to accurately describe all the homogeneous
normal and superfluid properties of the unitary Fermi gas.  The
\ASLDA\ functional thus provides means of using the quantitative
non-perturbative information about homogeneous phases to explore the
structure of inhomogeneous systems.  This has led to the prediction of
a non-trivial supersolid Larkin-Ovchinikov (\LO) phase~\cite{BF:2008}
of the polarized unitary Fermi gas.

\section{The Functional}
We choose to use an \ASLDA\ functional with the following form,
\begin{multline}
  E = \int \d^{3}{\vect{r}}\; \Biggl\{
  \mathcal{E}\bigl[n_{a,b}(\vect{r}),\tau_{a,b}(\vect{r}),
  \nu(\vect{r})\bigr] +\\
  + V_{\text{ext}}(\vect{r})\bigl[
  n_{a}(\vect{r}) + n_{b}(\vect{r})
  \bigr]
  + \text{sources}\ldots
  \Biggr\},
\end{multline}
where the (local) energy-density $\mathcal{E}$ is a function the densities.  We
construct the densities from the two-component Bogoliubov
(Nambu-Gor'kov) quasiparticle wavefunctions that satisfy the generalized Kohn-Sham equation
$\op{H}_{KS}\cdot\psi_{n} = E_{n}\psi_{n}$ (see below):
\begin{equation}
  \psi_{n}(\vect{r}) = \begin{pmatrix}
    u_{n}(\vect{r})\\
    v_{n}(\vect{r})
  \end{pmatrix}.
\end{equation}
The densities are constructed from these by summing over all
eigenstates $E_{n}$ weighted with the Fermi distribution function
$f_{\beta}(E)$, where $\beta = 1/T$.  (A small finite temperature can
help the initial part of the self-consistent calculations to converge,
see~\cite{Baran;Bulgac;Forbes;Hagen;Nazarewicz...:2008-05}.)
\begin{subequations}
  \label{eq:Densities}
  \begin{align}
    n_{a}(\vect{r}) =& \sum_{n}\abs{u_{n}(\vect{r})}^2f_{\beta}(E_{n}), \\
    n_{b}(\vect{r}) =& \sum_{n}\abs{v_{n}(\vect{r})}^2f_{\beta}(-E_{n}), \\
    \tau_{a}(\vect{r}) =& \sum_{n}\abs{\vect{\nabla}
      u_{n}(\vect{r})}^2f_{\beta}(E_{n}),\\ 
    \tau_{b}(\vect{r}) =& \sum_{n}\abs{\vect{\nabla}
      v_{n}(\vect{r})}^2f_{\beta}(-E_{n}),\\
    \nu(\vect{r}) =& \sum_{n} u_{n}(\vect{r})v_{n}^{*}(\vect{r})
    \frac{f_{\beta}(-E_{n}) - f_{\beta}(E_{n})}{2}.
  \end{align}
\end{subequations}
We express our functional in terms of these densities (from here on
we drop the explicit $\vect{r}$ dependence to simplify the notation)
\begin{subequations}
  \begin{multline}
    \label{eq:DF_ASLDA}
    \mathcal{E}\bigl[n_{a,b},\tau_{a,b},\nu\bigr] =\\
    =
    \alpha_{a}(n_{a},n_{b})\frac{\tau_{a}}{2} +
    \alpha_{b}(n_{a},n_{b})\frac{\tau_{b}}{2} +\\
    D(n_a,n_b) + g_{\text{eff}}(n_a,n_b)\nu^{*}\nu
  \end{multline}
  where $\alpha_{a,b}(\lambda n_{a},\lambda n_{b}) =
  \alpha(n_{a},n_{b})$ are the inverse effective masses, $D(\lambda
  n_{a},\lambda n_{b}) = \lambda^{5/3}D(n_{a},n_{b})$ contains the
  density dependent portion of the functional and $C(\lambda
  n_{a},\lambda n_{b}) = \lambda^{1/3}C(n_{a},n_{b})$ is an
  appropriate density dependent inverse effective coupling, with
  regulator $\Lambda$:
  \begin{equation}
    \frac{1}{g_{\text{eff}}} = C(n_{a},n_{b}) - \Lambda.
  \end{equation}
  The divergences appear in both the pairing and kinetic terms such
  that $\Delta = -g_{\text{eff}}\nu$ and the combination
  \begin{equation}
    \alpha_{a}\frac{\tau_{a}}{2} + 
    \alpha_{b}\frac{\tau_{b}}{2} - g_{\text{eff}}\nu^{*}\nu
  \end{equation}
  are finite.  Thus, the energy functional may be expressed in terms
  of finite combinations
  \begin{equation}
    \mathcal{E} =
    \left[
      \alpha_{a}\frac{\tau_{a}}{2} +
      \alpha_{b}\frac{\tau_{b}}{2} 
      - \Delta^{*}\nu
    \right]
    + D
  \end{equation}
  where $\Delta^{*}\nu = \nu^{*}\Delta = 
  -g_{\text{eff}}\nu^{*}\nu$.
\end{subequations}
One recovers a typical Kohn-Sham functional by setting
$\alpha_{a,b}\equiv 1$ and $\Delta \equiv 0$ ($\nu\equiv 0)$.
Regularization is required to describe superfluids ($\Delta \neq 0$
and $\nu \neq 0$) because the anomalous density $\nu$ and the kinetic
energy densities $\tau_{a,b}$ diverge.  This regulator dependence is
removed when one considers the proper combinations of terms described
above.  Allowing the effective mass to deviate $\alpha \neq 1$ in the
\SLDA\ description of the trapped fermions improved the agreement with
Monte-Carlo (\MC) results~\cite{Bulgac:2007a,Chang;Bertsch:2007-08,Blume;Stecher;Greene:2007-12,Stecher;Greene;Blume:2008-12}, and can be constrained
by non-perturbative results, so we keep this generalization.  The
coefficient functions $\alpha_{a}(n_{a},n_{b})$,
$\alpha_{b}(n_{a},n_{b})$, $C(n_{a},n_{b})$, and $D(n_{a},n_{b})$ are
chosen to match the thermodynamic properties of the system in the
homogeneous limit.  The lack of scales in the unitary limit further
constrains the forms of these coefficients to be homogeneous functions
of specific degree in the densities $n_{a,b}$.

We established this form of the \ASLDA\ functional using the following
guiding principles:
\begin{description}
\item[Gradient Expansion:] We start by formulating a functional
  applicable to slowly varying systems.  To this end, we have
  neglected gradient terms.  Indeed, the symmetric \SLDA\ functional
  has had remarkable success \emph{without any} gradient corrections
  \cite{Bulgac:2007a}, so we assume that this property holds for
  asymmetric systems and start with only local terms.  An important
  future direction will be to quantify the effects of gradient
  corrections, and to extract the coefficients of the leading
  gradient terms.

  Note that the success of the \SLDA\ implies that the subset of
  gradient corrections included implicitly through the standard
  kinetic terms provides the dominant gradient contribution.  These
  corrections are also included in the \ASLDA. 

  To justify the omission of gradient terms (beyond those contained in
  $\tau_{a,b}$), consider the lowest order gradient correction to the
  \SLDA\ functional: this is of the form $\abs{\vect{\nabla}n}^2/n$
  and would lead to a correction in the total energy of a system of
  $N$ harmonically trapped fermions that scales as $N^{2/3}$.  The
  extremely good agreement between the \SLDA\ functional in
  Ref.~\cite{Bulgac:2007a} and the \emph{ab initio} results of
  Refs.~\cite{Chang;Bertsch:2007-08,Blume;Stecher;Greene:2007-12,Stecher;Greene;Blume:2008-12}
  indicate that the coefficient of this correction must be extremely
  small.  One can also show that, in the dilute limit, the strength of
  the gradient terms is controlled by the size of the length scales
  such as the effective range and \textit{p}-wave scattering length of
  the
  interaction~\cite{Bhattacharyya;Furnstahl:2005-10,Bhattacharyya;Furnstahl:2005-12}.
  In the unitary Fermi gas, both of these vanish.

  In addition, the leading gradient corrections that could appear away
  from the symmetric limit $x=n_{b}/n_{a} = 1$ are suppressed by
  $(1-x)^2$.  Thus, we expect them to have little effect on phases
  close to the fully paired \SF\ state, for example, the \LO\ phase
  discussed in~\cite{BF:2008}.

  Finally, Galilean invariance requires that if $\alpha_{a,b}\neq 1$
  then $\tau_{a,b}$ be replaced by $\tau_{a,b} -
  \vect{p}_{a,b}^2/n_{a,b}$ where $\vect{p}_{a,b}$ is the local
  current density~\cite{Bulgac:2007a}.  This correction is only
  required to discuss states that break time-reversal invariance
  (i.e. that contain currents).
\item[Simplicity:] Due to the fermion sign problem, there are very few
  reliable calculations of properties in the polarized regime, thus
  there is not enough data to properly constrain a fully general
  functional.  Monte-Carlo simulations, however, have provided
  reasonably constraints on the form of the normal state energy
  density, thus we allow for full generality in terms of the
  functional dependence on the densities, but restrict the general
  dependence on the anomalous and kinetic densities.  The most general
  form of local function would allow the unknowns to depend on all of
  the dimensionless combinations of densities, including for example,
  an arbitrary dependence on the dimensionless regulator-invariant
  combination $[\alpha_{a}\tau_{a}/2 + \alpha_{b}\tau_{b}/2 -
  \Delta^{*}\nu]/n^{5/3}$.  Presently, we see no need for this
  added complication.
\item[Quasiparticle Dispersion:] Monte-Carlo calculations about the
  symmetric phase have suggested that the low-temperature
  quasiparticle dispersions are well described by the \BCS\ form
  \begin{equation}
    \label{eq:Eqp}
    E_{k} = \sqrt{\left[\alpha\frac{k^2}{2} + (U-\mu)\right]^{2}
      + \abs{\Delta}^{2}}
  \end{equation}
  where $\alpha$, $(U-\mu)$, and $\Delta$ are effective parameters.
  This form of dispersion follows from introducing the anomalous
  pairing density $\nu$ through the quadratic form $\nu^{*}\nu$ along
  with canonical kinetic terms.
\item[Decoupling of Paired and Normal States:] In our actual
  formulation, we also neglect the general density dependence of the
  pairing interaction $C(n_{a},n_{b})$, replacing this by the same
  term with a single constant used in the \SLDA.  The justification of
  this is two-fold: 1) The \SLDA\ worked very well, even including
  small polarizations. 2) The energy of interaction of the normal
  state is well approximated by considering only the parameters of the
  \SLDA\ without any additional density dependence.

  Together these suggest that the physics of the superfluid state
  somewhat decouples from the physics of the normal state, allowing
  one to characterize the fully paired superfluid independently of the
  normal state.

  This implies a qualitative ansatz of this functional: that the
  structure of the polarized phases arises from the competition
  between the fully paired superfluid physics and the interacting
  normal state physics.  In principle, it is possible that some
  qualitatively new description is required to properly account for
  the structure of polarized fermionic matter -- for example to
  describe the appearance of \textit{p}-wave pairing at large
  polarizations~\cite{Bulgac:2006gh} -- but to fix such a description
  will require high precision \emph{ab initio} calculations and/or
  experiments that have not yet been realized.
\end{description}
Varying this functional with respect to the quasi-particle
wavefunctions and occupation numbers gives the generalized Kohn-Sham
equation $\op{H}_{KS}\cdot\psi_{n} = E_{n}\psi_{n}$, where
\begin{subequations}
  \begin{equation}
    \op{H}_{KS} = 
    \begin{pmatrix} 
      \op{K}_{a} - \mu_{a} + U_{a} & \Delta^{*}\\
      \Delta & -\op{K}_{b} + \mu_{b} - U_{b}
    \end{pmatrix},
  \end{equation}
  and the kinetic and potential operators are
  \begin{align}
    \op{K}_{a}u &= -\frac{1}{2}\vect{\nabla}\cdot(\alpha_{a}\vect{\nabla}u)\\
    U_{a} &= \pdiff{\alpha_{-}}{n_{a}}
    \frac{\tau_{-}}{2}
    +\pdiff{D}{n_{a}}
    -\pdiff{C}{n_{a}}\abs{\Delta}^2+\nonumber\\
    & \quad
    +\pdiff{\ln\alpha_{+}}{n_{a}}\left[
      \left(
        \alpha_{+}
        \frac{\tau_{+}}{2}
        -\Delta^{*}\nu
      \right)
      -  
      C\abs{\Delta}^2
    \right],
  \end{align}
  with the notations
  \begin{align}
    \alpha_{\pm} &= \frac{\alpha_{a}\pm\alpha_{b}}{2}, &
    \tau_{\pm} &= \tau_{a} \pm \tau_{b}, &
    n_{\pm} &= n_{a} \pm n_{b}.
  \end{align}
  The form of these operators for species $b$ are obtained by
  interchanging $a \leftrightarrow b$.  Note that the terms have been
  grouped so that the ultraviolet (\textsc{uv}) divergences arising
  from the local form of the anomalous density $\nu$ cancel in the
  last term (see Refs. \cite{Bulgac:2007a,BY:2003} for details).  All
  other terms are finite.
\end{subequations}
\section{Fitting the Functional}
To further specify the functional, we must fix the forms of the
functions $\alpha_{a,b}(n_{a},n_{b})$, $C(n_{a},n_{b})$, and
$D(n_{a},n_{b})$.  To do this, we characterize the
thermodynamic properties of the system, which are fortunately quite
tightly constrained~\cite{Bulgac:2006cv}, and have both
calculational~\cite{Bulgac:2006cv,Chevy:2006b,LRGS:2006,CRLC:2007}
and experimental~\cite{shin-2008,Carlson;Reddy:2008-04} verification.

The form of the functional allows us to consider two different species
(for example, with different masses), but we are interested in systems
with two identical species.  The functional must thus exhibit the
discrete symmetry $a\leftrightarrow b$.  This constrains the form of
the functions $\alpha_{a,b}$, $D$ and $C$:
\begin{subequations}
  \label{eq:symmetry}
  \begin{align}
    \alpha_{a}(n_{a},n_{b}) &= \alpha_{b}(n_{b},n_{a})\\
    D(n_{a},n_{b}) &= D(n_{b},n_{a})\\
    C(n_{a},n_{b}) &= C(n_{b},n_{a}).
  \end{align}
\end{subequations}
Dimensional analysis determines the overall scaling of the functions,
and we may fully parametrize the functional with three dimensionless
functions $\alpha(x)$, $b(x)$, and $\gamma(x)$ of the asymmetry
$x=n_{b}/n_{a} \in [0,1]$, with the complementary region determined by
symmetries~(\ref{eq:symmetry}):
\begin{subequations}
  \label{eq:functions_x}
  \begin{align}
    \alpha_{a}(n_{a},n_{b}) &= \alpha(x),\\
    \alpha_{b}(n_{a},n_{b}) &= \alpha(1/x),\\
    D(n_{a},n_{b}) &=
    \frac{(3\pi^2)^{5/3}(n_{a}+n_{b})^{5/3}}{10\pi^2}\beta(x),\\
    C(n_{a},n_{b}) &= \frac{(n_{a}+n_{b})^{1/3}}{\gamma(x)}.
  \end{align}
\end{subequations}
As a technical note, we ensure that our parametrization is smooth at
$x=1$ by letting $\alpha(x)$ be smooth over $[0,\infty)$, and by
forming smooth even functions over the variable $z\in[-1,1]$ where
$\ln{x} = c\tanh^{-1}{z}$: thus $\tilde{\beta}(z) =
\beta\bigl(x(z)\bigr)$ and $\tilde{\gamma}(z) =
\gamma\bigl(x(z)\bigr)$ are smooth at $x=1$ if and only if
$\tilde{\beta}(z)$ and $\tilde{\gamma}(z)$ are smooth even functions.
The choice of the parameter $c$ is made so that the interpolations of
the Monte-Carlo data are well behaved.

To determine these dimensionless functions, we match the functional to
the Monte-Carlo calculations of pure and homogeneous thermodynamic phases.
These phases possess no gradients, and so solving the density
functional for homogeneous matter is equivalent to performing a simple
Thomas Fermi type calculation with the added complication that the
parameters: $m_{a,b}$, etc. depend on the densities, which must be
determined self-consistently.  This gives rise to a set of non-linear
set of equations that can be fairly easily solved.

We start with the homogeneous fully paired superfluid phase.  This is
described by the three numbers $\alpha=\alpha(1)$,
$\beta=\beta(1)$, and $\gamma=\gamma(1)$ which may be
extracted by fitting the quasi-particle dispersion relationship and
the energy.  As discussed earlier, we simplify the functional
dependence of the function $\gamma(x) = \gamma$ by simply keeping this
constant.  The remaining functional forms $\alpha(x)$ and $\beta(x)$
are determined by fitting the energy of the homogeneous normal state
to Monte-Carlo results.  This completely specifies the functional in a
unique manner as we shall now describe.

\subsection{Symmetric Superfluid Properties:}
As suggested in~\cite{Bulgac:2007a}, by considering the calculated
properties of the fully paired symmetric superfluid, one may determine
the values of the functions $\alpha(1)$, $\beta(1)$, and $\gamma(1)$
at the point $x=n_{b}/n_{a}=1$.  We start by taking
\begin{equation}
  n_{a} = n_{b} = n = \frac{n_{+}}{2}
\end{equation}
where $n_{+} = n_{a} + n_{b} = 2n$ is the total density.  Dimensional
analysis determines the following forms of the derivatives (evaluated at
$n_{a} = n_{b} = n$):
\newcommand{\args}{(n,n)}
\begin{subequations}
  \begin{align}
    \pdiff{\alpha_{+}}{n_{a}}
    = \pdiff{\alpha_{+}}{n_{b}} &= 0,\\
    \pdiff{C}{n_{a}}
    = \pdiff{C}{n_{b}}
    &= \frac{C}{3n},\\
    \pdiff{D}{n_{a}}
    = \pdiff{D}{n_{b}} &= \frac{5 D}{3n}.
  \end{align}
\end{subequations}
Thus, the effective potentials $U_{a} = U_{b} = U$ have
the following simplified form,
\begin{equation}
  U\args = \frac{5 D\args}{3n} -
  \frac{C\args}{3n}\Delta^{*}\Delta,
\end{equation}
and one may take a linear combination of the kinetic terms to obtain 
\begin{equation}
  \op{K}_{+} = \frac{\op{K}_{a} + \op{K}_{b}}{2} = 
  -\frac{\alpha_{+}\vect{\nabla}^2}{2}.
\end{equation}
Thus, the symmetric phase depends only on three parameters
$\alpha = \alpha(1)$, $\beta = \beta(1)$, and $\gamma = \gamma(1)$ via:
\begin{subequations}
  \begin{align}
    \alpha_{+}\args &= \alpha(1),\\
    C\args &= \frac{n_{+}^{1/3}}{\gamma(1)},\\
    D\args &= \beta(1)\;\frac{(3\pi^2n_{+})^{5/3}}{10\pi^2}
    = \beta(1)\;\mathcal{E}_{FG}.
  \end{align}
\end{subequations}
For a given inverse effective mass $\alpha$, the other two parameters
$\gamma$ and $\beta$ may be fit by requiring that the energy and
spectral gap satisfy
\begin{subequations}
  \label{eq:SymmConditions}
  \begin{align}
    \mathcal{E}_{SF} = \mathcal{E}\args &= \xi \mathcal{E}_{FG} = 
    \xi\frac{(3\pi^2n_{+})^{5/3}}{10\pi^2},\\
    \Delta &= \eta\epsilon_{F} 
    = \eta\frac{(3\pi^2n_{+})^{2/3}}{2},
  \end{align}
\end{subequations}
where
\begin{subequations}
  \begin{align}
    k_{F} &= (3\pi^2n_{+})^{1/3},\\
    \epsilon_{F} &= \frac{k_{F}^2}{2} 
    = \frac{(3\pi^2n_{+})^{2/3}}{2},\\
    \mathcal{E}_{FG} &= 2\frac{k_{F}^5}{20\pi^2}
    = \frac{(3\pi^2n_{+})^{5/3}}{10\pi^2}
    = \frac{3}{5}n_{+}\epsilon_{F}.
  \end{align}
\end{subequations}
The $T=0$ symmetric state is characterized by the integrals
\begin{subequations}
  \begin{align}
    n_{+} &= \int\frac{\d^{3}\vect{k}}{(2\pi)^3}\;\left[1 - \frac{\epsilon_{k}}{E_{k}}\right],
    \label{eq:n_sym}\\
    \mathcal{E}_{SF} &= \int\frac{\d^{3}{\vect{k}}}{(2\pi)^2}\left\{
      \alpha\frac{k^2}{2}\left[1 - \frac{\epsilon_{k}}{E_{k}}\right]
      - \frac{\abs{\Delta}^2}{2E_{k}}
    \right\} + \beta\mathcal{E}_{FG},
    \label{eq:ESF_sym}\\
    C &= \frac{n_{+}^{1/3}}{\gamma} = -
    \int\frac{\d^{3}\vect{k}}{(2\pi)^3}\
    \left[
      \frac{1}{2E_{k}}
      -\frac{1}{\alpha k^2}
    \right],\label{eq:C_sym}
  \end{align}
  where 
  \begin{align}
    \epsilon_{k} &= \alpha\frac{k^2}{2} + (U - \mu), \\
    E_{k} &= \sqrt{\epsilon_{k}^2 + \abs{\Delta}^{2}} .
  \end{align}
\end{subequations}
Given fixed values of $\alpha$, $\xi$, and $\eta$, we proceed
as follows: 1) Choosing the density $n$ so that $\epsilon_{F}=1$: This fixes
the scale and determines $\Delta$, 2) Use equation~(\ref{eq:n_sym}) to
solve for the combination $U-\mu$ that appears on the right hand side
through $\epsilon_{k}/E_{k}$, 3) Use equation~(\ref{eq:ESF_sym}) and $\xi$ to solve
for $\beta$, and 4) Use equation~(\ref{eq:C_sym}) to solve for $\gamma$.

The parameters $\xi$ and $\eta = \Delta/\epsilon_{F}$ have been
measured by several Monte-Carlo
techniques~\cite{CCPS:2003,Carlson:2005kg,Carlson:private_comm2,BDMW:2008,BDM:2008}.
We take the following values in our
estimates~\cite{Carlson:2005kg,Carlson:private_comm2}:
\begin{align}
  \label{eq:xi_Delta0}
  \xi &= \frac{\mathcal{E}\args}{\mathcal{E}_{FG}\args} = 0.40(1),&
  \eta &= \frac{\Delta}{\epsilon_{F}} = 0.504(24).
\end{align}
Note that it was incorrectly stated in~\cite{Bulgac:2007a} that
$\alpha_{x=1}$ could also be determined through the condition that the
chemical potential satisfy $\mu = \xi\epsilon_{F}$, but a careful
examination shows that this is implied by~(\ref{eq:SymmConditions}).
It is also clear in DFT's developed
perturbatively~\cite{Bhattacharyya;Furnstahl:2005-10,Bhattacharyya;Furnstahl:2005-12}
that the effective mass is arbitrary.  In order to fix the effective
mass, we match the quasi-particle dispersion relationship as
determined from the Monte-Carlo results~\cite{Carlson:2005kg}.  The
dispersion relationships within our density functional have the form
\begin{equation}
  \label{eq:qp_disp}
  \frac{E_{k}}{\epsilon_{F}} = 
  \sqrt{\left[\frac{\alpha}{2}\frac{k^2}{k_{F}^2} 
      + \left(\frac{U}{\epsilon_{F}} - \xi\right)
    \right]^2 
    + \left(\frac{\Delta}{\epsilon_{F}}\right)^2}.
\end{equation}
Note that the combination $U/\epsilon_{F} - \xi$ is fixed from~(\ref{eq:n_sym})
and depends only on $\Delta/\epsilon_{F}$, so the quasiparticle dispersion
relation is sensitive only to $\Delta/\epsilon_{F}$ and the effective mass.  We
add the value (\ref{eq:xi_Delta0}) as an additional data-point in
the fit and perform a non-linear least-squares fit.

The fit to the Carlson-Reddy is shown in Fig.~\ref{fig:qp1} and gives
the following parameter values:\footnote{Here we have performed a
  simple three-parameter non-linear least-squares fit: this has a
  quality factor $Q=0.52$ which is quite good.  A different analysis
  would hold $\Delta$ and $U$ fixed to the properly determined values,
  but the method here is consistent since the errors are of the same
  magnitude.}
\begin{subequations}
  \begin{align}
    \alpha_{x=1} = m_{\text{eff}}^{-1}/m^{-1} &= \phantom{-}1.094(17),\\
    \beta_{x=1} &= -0.526(18), \\
    \gamma^{-1} &= -0.0907(77)\\
    \bar{\beta} = U/\epsilon_{F} &= -0.491(18),\\
    \eta = \Delta/\epsilon_{F} &= \phantom{-}0.493(12),
    \label{eq:m_Delta0}\\
    \xi_{N} = \alpha + \beta &= \phantom{-}0.567(24).
  \end{align}
\end{subequations}
where
\begin{equation} % Checked.
  \frac{U}{\epsilon_{F}} = 
  \beta -
  \frac{(3\pi^2)^{2/3}}{6\gamma}\left(\frac{\Delta}{\epsilon_{F}}\right)^2.
\end{equation}
In principle, one should use some form of \emph{ab initio} calculation
or experimental measurement for polarized systems to determine the
dependence of the parameters $\alpha$, $\beta$, and $\gamma$ on the
polarization $x=n_{b}/n_{a}$.  Unfortunately, the fermion sign problem
has made this difficult and there is presently insufficient quality
data to perform such a fit.  Instead, we make the approximation that
\begin{equation}
  \gamma(x) = \gamma(1) = \text{const}.
\end{equation}

\subsection{Normal State}
The remaining functional forms for $\alpha(x)$ and $\beta(x)$ can be
extracted from properties of the homogeneous normal state at $T=0$ by
explicitly setting the anomalous density $\nu = 0$.  This metastable
state may be explored by choosing a suitable nodal approximation in
the fixed-node Monte-Carlo (\textsc{fn}-\MC) calculations.  Such a
restriction seems to remove most of the superfluid correlations from
the results, but we suspect that one cannot completely remove all
contamination from the superfluid state for small polarizations.  For
this reason, we have only included the data for large polarizations $x
= n_{b}/n_{a} < 0.5$ in our fitting of the parameters for the
functional (see Fig.~\ref{fig:giorgini} and Tab.~\ref{tab:G_interp}).

The energy-density for the normal phase of homogeneous matter has the
form
\begin{align}
  \mathcal{E}[n_{a},n_{b}] &=
  \frac{\alpha_{a}(6\pi^2n_{a})^{5/3}}{20\pi^2}
  +
  \frac{\alpha_{b}(6\pi^2n_{b})^{5/3}}{20\pi^2} 
  + D \nonumber\\
  &= \frac{(6\pi^2)^{5/3}(n_{a}+n_{b})^{5/3}}{20\pi^2}G(x)
\end{align}
where
\begin{align}
  \label{eq:G_x}
  G(x) &= \frac{\alpha(x)}{(1+x)^{5/3}} + 
  \frac{\alpha(x^{-1})}{(1+x^{-1})^{5/3}} + 2^{-2/3}\beta(x),\nonumber\\
  &= \frac{1}{(1+x)^{5/3}}g^{5/3}(x).
\end{align}

From this relationship, one can uniquely determine the functional
form for $\beta(x)$ given a form for the inverse effective mass
$\alpha(x)$, which we shall construct below, and the normal state
energy density $g(x)$, which has been well-constrained by Monte-Carlo
data~\cite{LRGS:2006} (see Fig.~\ref{fig:giorgini}).

\begin{figure}[b]
  \begin{center}
    \psfrag{z}{$z$}
    \psfrag{alpha_z}{\hspace{-1em}$\alpha(z)= m_{\text{eff}}^{-1}$}
    \psfrag{0.9}{\footnotesize 0.9}
    \psfrag{1.1}{\footnotesize 1.1}
    \psfrag{1.2}{\footnotesize 1.2}
    \psfrag{-1.0}{\footnotesize -1.0}
    \psfrag{-0.5}{\footnotesize -0.5}
    \psfrag{0.0}{\footnotesize 0.0}
    \psfrag{0.5}{\footnotesize 0.5}
    \psfrag{1.0}{\footnotesize 1.0}
  \includegraphics[width=\columnwidth]{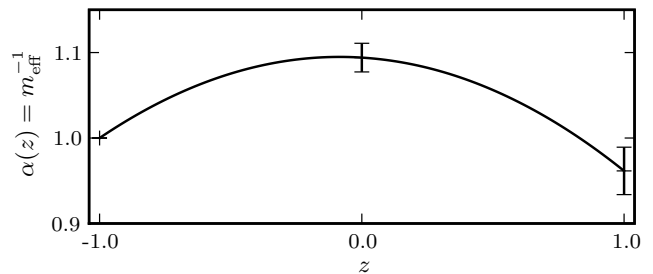}
    \caption{\label{fig:alpha_z}
      Inverse effective mass $\alpha=m_{\text{eff}}^{-1}$ as a function of
      $z=\tanh\tfrac{1}{c}\ln{x}$ with $c=1$.  The function is a
      parabolic fit through the three data-points.
    }
  \end{center}
\end{figure}
\begin{figure}[b]
  \begin{center}
    \psfrag{x}{\hspace{-7em}$x$ (lower curves), $1/x$ (upper curves)}
    \psfrag{m_eff=1/alpha}{$m_{\text{eff}}/m = \alpha^{-1}$}
    \psfrag{c=0.5}{\footnotesize $c=1/2$}
    \psfrag{c=1.0}{\footnotesize $c=1$}
    \psfrag{c=2.0}{\footnotesize $c=2$}
    \psfrag{0.9}{\footnotesize 0.9}
    \psfrag{1.0}{\footnotesize 1.0}
    \psfrag{1.1}{\footnotesize 1.1}
    \psfrag{0.0}{\footnotesize 0.0}
    \psfrag{0.2}{\footnotesize 0.2}
    \psfrag{0.4}{\footnotesize 0.4}
    \psfrag{0.6}{\footnotesize 0.6}
    \psfrag{0.8}{\footnotesize 0.8}
    \psfrag{1.0}{\footnotesize 1.0}
    \includegraphics[width=\columnwidth]{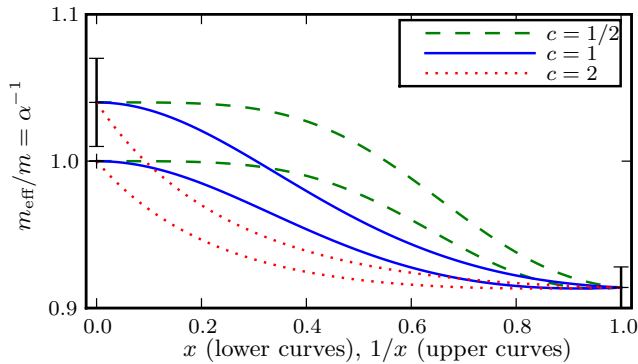}
    \caption{\label{fig:m_x_c} (Color online)
      Effective mass as a function of
      $x=n_{b}/n_{a}$ for several values of $c$: $c=1/2$ -- dashed
      green, $c=1$ -- solid blue, $c=2$ -- dotted red.  The upper
      curves represent the extension of the curves to $x>1$ with
      abscissa $1/x$.  Since the effective mass is not likely to
      change rapidly near $x=0$ (especially for the majority
      component), the parameter $c$ is probably best kept near $1$ and
      we use $c=1$.}
  \end{center}
\end{figure}

\subsection{Effective Mass Parametrization: $\alpha(x)$}

Several different Monte-Carlo calculations have constrained the
effective ``polaron'' mass of a single spin-down fermion in a sea of
spin-up fermions, $m_{0}\approx
1.04(3)m$~\cite{LRGS:2006,CRLC:2007,PS:2008}, and the
effective mass of the particles in the background gas will not be
altered in the limit of extreme polarization.  Thus, the endpoints of
the function $\alpha(x)$ are constrained: $\alpha(0) = 1$,
$\alpha(\infty) = 0.96(3)$.  A third point is obtained from the
effective mass $\alpha(1) = 1.09(2)$ in the fully paired superfluid
phase~(\ref{eq:m_Delta0}).  We use these to provide a smooth
interpolation via the parametrization
\begin{equation}
  \label{eq:z_c}
  z_{c} = \tanh\frac{\ln x}{c} 
\end{equation}
as shown in Fig.~\ref{fig:alpha_z}.

We use the variable $z_{c}$~(\ref{eq:z_c}) here so that the
interpolation is over a finite range $z_{c}\in[-1,1]$.  The parameter
$c$ gives us some control over the shape of the resulting curve as
demonstrated in Fig.~\ref{fig:m_x_c}.  Note that for $c<2$,
$\left.\d{z_{c}}/\d{x}\right|_{x=0} = 0$, hence the function
$\alpha(x)$ will be flat at $x=0$.  The addition of a few spin-down
particles should not affect the mass of the spin-up particle: most of
the change should occur when one approaches equal densities.  There is
not really sufficient information to further characterize this
parametrization, but the effective mass does not vary much, so we do
not expect this to be a significant source of error.

Including density dependent inverse masses $\alpha_{a,b}(n_{a},n_{b})$
can be of quantitative importance, but does not significantly alter
the qualitative aspects of the \ASLDA\ (such as presented
in~\cite{BF:2008}).  Thus, to obtain qualitative results, it can be a
good first approximation to simply use the functional with
$\alpha_{a,b}(n_{a},n_{b}) = 1$.

\subsection{``Hartree'' Energy : $\beta(x)$.}
\begin{figure}[thb]
  \begin{center}
    \psfrag{z}{$z$}
    \psfrag{G}{$\tilde{G}(z)$}
    \psfrag{cG=1.0}{\footnotesize $c=1$}
    \psfrag{cG=2.0}{\footnotesize $c=2$}
    \psfrag{0.3}{\footnotesize 0.3}
    \psfrag{0.7}{\footnotesize 0.7}
    \psfrag{1.1}{\footnotesize 1.1}
    \psfrag{-1.0}{\footnotesize -1.0}
    \psfrag{-0.5}{\footnotesize -0.5}
    \psfrag{0.0}{\footnotesize 0.0}
    \psfrag{0.5}{\footnotesize 0.5}
    \psfrag{1.0}{\footnotesize 1.0}
    \includegraphics[width=\columnwidth]{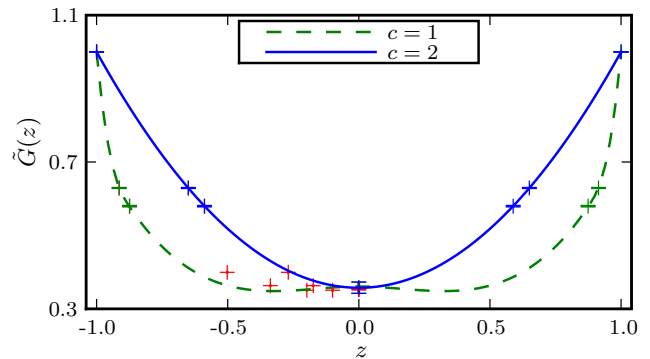}
    \caption{\label{fig:G_z_c} (Color online) The function
      $\tilde{G}(z)$ defined in~(\ref{eq:G_x}) plotted for various
      values of $c$ used to define the interpolation through the
      coordinate $z_{c}$~(\ref{eq:z_c}): $c=1$ -- dashed green, $c=2$
      -- solid blue.  See Fig.~\ref{fig:G_x_c} which better emphasizes
      how this value is a better fit.  We omitted the red points
      from our interpolation as discussed in the text.}
  \end{center}
\end{figure}

\begin{figure}[thb]
  \begin{center}
    \psfrag{f(x)=g^{5/3}(x)}{$f(x)=g^{5/3}(x)$}
    \psfrag{x=n_b/n_a}{$x=n_{b}/n_{a}$}
    \psfrag{cG=0.5}{\footnotesize $c=1/2$}
    \psfrag{cG=1.0}{\footnotesize $c=1$}
    \psfrag{cG=2.0}{\footnotesize $c=2$}
    \psfrag{0.8}{\footnotesize 0.8}
    \psfrag{0.9}{\footnotesize 0.9}
    \psfrag{1.0}{\footnotesize 1.0}
    \psfrag{1.1}{\footnotesize 1.1}
    \psfrag{0.0}{\footnotesize 0.0}
    \psfrag{0.2}{\footnotesize 0.2}
    \psfrag{0.4}{\footnotesize 0.4}
    \psfrag{0.6}{\footnotesize 0.6}
    \includegraphics[width=\columnwidth]{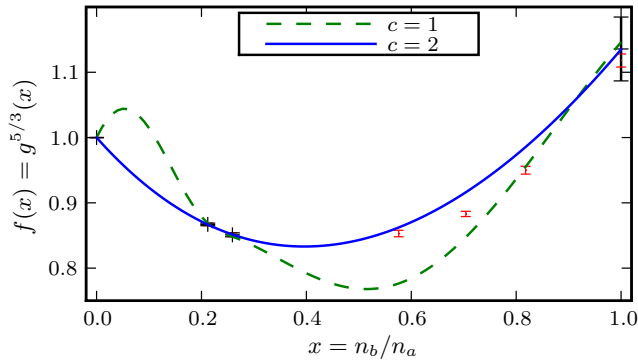}
    \caption{\label{fig:G_x_c} (Color online) The function
      $f(x)=g^{5/3}(x)$.  Both function plotted for various values of
      $c$ used to define the interpolation through the coordinate
      $z_{c}$~(\ref{eq:z_c}): $c=1$ -- dashed green, $c=2$ -- solid
      blue.  It can be clearly seen here that the value of $c=2$ is
      preferred for a smooth interpolation.}
  \end{center}
\end{figure}

To finish the parametrization, we must provide an interpolation of the
function $g(x)$.  We provide an interpolation for the function
$G(x)$~(\ref{eq:G_x}) rather than directly for $\beta(x)$ or $g(x)$
because: 1) It is finite everywhere and, 2) such an interpolation is
independent of the inverse effective mass function $\alpha(x)$.
Again, we use the parametrization~(\ref{eq:z_c}) with $z_{c} \in
[-1,1]$ with the assumption that $G(x)$ is smooth at $x=1$, which
requires that the interpolated function $\tilde{G}(z) =
G\bigl(x(z)\bigr)$ is smooth and even.  This is quite easy to do (one
can always just explicitly make the interpolation even $[\tilde{G}(z)
+ \tilde{G}(-z)]/2$).  The assumption of smoothness also gives a
non-trivial constraint on the data which should admit a smooth
interpolation at $z_{c}=0$.  The resulting function is shown in
Fig.~\ref{fig:G_z_c} for several values of $c$.  Here the value of
$c=2$ is best because $\d{z_{c}}/\d{x}|_{x=0} = 2$ is finite,
preserving the structure of the interpolation.  (The other values of
$c$ drastically affect the slope of the interpolated $g(x)$ without
any physical motivation.)  The preference for $c=2$ is most evident in
Fig.~\ref{fig:G_x_c}.


\section{LOFF}
Finally we include the derivation of a simple relationship between the
average pressure and energy density of a \LOFF\ state with period $L$ imposed
by boundary conditions.  The physical solution will have a
spontaneously determined length scale $L$, but our calculation must
impose this through boundary conditions.  To model these states, we
use a periodic basis, summing over transverse and Bloch momenta.  One
must then adjust $L$ to maximize the average pressure.

Consider the form of the average pressure at unitarity.  The lack of
scales implies
\begin{equation}
  s^{5}\mathcal{P}(s^{-2}\mu_{a},s^{-2}\mu_{b},sL) 
  = \mathcal{P}(\mu_{a},\mu_{b},L).
\end{equation}
Differentiating and using the thermodynamic relationship $n_{a,b} =
\partial\mathcal{P}/\partial\mu_{a,b}$, we obtain:
\begin{equation}
  5\mathcal{P} - 2\mu_{a}n_{a} - 2\mu_{b}n_{b} + L \pdiff{\mathcal{P}}{L} = 0.
\end{equation}
Coupled with the thermodynamic relationship $\mathcal{P} =
\mu_{a}n_{a} + \mu_{b}n_{b} - \mathcal{E}$
we have
\begin{equation}
  X = L\pdiff{\mathcal{P}}{L} = 2\mathcal{E} - 3\mathcal{P}.
\end{equation}
When the pressure is maximized, this quantity $X=0$.
Note that this relationship is derived solely on dimensional grounds.
\section{Interpolations:}
The functional~(\ref{eq:DF_ASLDA}) is completely described by the
three functions $\alpha(x)$, $\beta(x)$, and $\gamma(x)$ through
equations~(\ref{eq:functions_x}).  These functions are defined by the
functions $\tilde{\alpha}(z)$ and $\tilde{G}(z)$ that interpolate the
data in tables~\ref{tab:alpha_interp} and \ref{tab:G_interp}.  The
resulting weighted cubic-spline interpolations are shown in
Fig.~\ref{fig:alpha_z} and Fig.~\ref{fig:G_z_c}) respectively.
\begin{align*}
  \alpha(x) &= \tilde{\alpha}\left(\tanh\ln{x}\right),\\
  \beta(x) &= 2^{2/3}\tilde{G}\left(\tanh\frac{\ln{x}}{2}\right) +\\
  &\qquad - \frac{\alpha(x)}{(1+x)^{5/3}} 
  - \frac{\alpha(x^{-1})}{(1+x^{-1})^{5/3}},\\ 
  \gamma(x) &= \gamma.
\end{align*}
\begin{table}[htb]
  \caption{\label{tab:alpha_interp}
    Interpolation points for the function $\tilde{\alpha}(z)$.}
  \begin{ruledtabular}
    \begin{tabular}[t]{D{.}{.}{0}D{.}{.}{7}}
      \multicolumn{1}{c}{$z$} & 
      \multicolumn{1}{c}{$\tilde{\alpha}$} \\
      \hline
      -1 & 1.000(00)\\
      0 & 1.094(17)\\
      1 & 0.962(28)
    \end{tabular}
  \end{ruledtabular}
\end{table}
\begin{table}[htb]
  \caption{\label{tab:G_interp}
    Interpolation points for the function $\tilde{G}(z)$.}
  \begin{ruledtabular}
    \begin{tabular}[t]{D{.}{.}{4}D{.}{.}{8}}
      \multicolumn{1}{c}{$z$} & 
      \multicolumn{1}{c}{$\tilde{G}$} \\
      \hline
      -1.0000 & 1.0000(00)\\
      -0.6502 & 0.6293(15)\\
      -0.5886 & 0.5797(20)\\
      0.0000 & 0.3577(15)\\
      0.5886 & 0.5797(20)\\
      0.6502 & 0.6293(15)\\
      1.0000 & 1.0000(00)\\
    \end{tabular}
  \end{ruledtabular}
\end{table}
\acknowledgments We acknowledge the US Department of Energy for
support under Grants No. DE-FG02-97ER41014 and DE-FC02-07ER41457.

\clearpage
%\bibliographystyle{h-physrev3}
%\bibliography{master}

\end{document}